\documentstyle[12pt,aaspp4]{article}
  
\newcommand {\go} {\mathrel{\hbox{\rlap{\lower.55ex \hbox {$\sim$}}
        \kern-.3em \raise.4ex \hbox{$>$}}}}
\newcommand {\lo} {\mathrel{\hbox{\rlap{\lower.55ex \hbox {$\sim$}}
        \kern-.3em \raise.4ex \hbox{$<$}}}}
\newcommand {\be} {\begin{equation}}
\newcommand {\ee} {\end{equation}}

\begin{document}

\title{X-RAY EMISSION FROM THE QUASAR PKS 1127-145: 
   COMPTONIZED IR PHOTONS ON PARSEC SCALES}
\author{Micha{\l} B{\l}a\.zejowski, Aneta Siemiginowska} 
\affil{Harvard-Smithsonian Center for Astrophysics, 60 Garden Street,
Cambridge, MA 02138, USA, mblazejowski@cfa.harvard.edu, aneta@head-cfa.harvard.edu} 
\author{Marek Sikora, Rafa{\l} Moderski}
\affil{Nicolaus Copernicus Astronomical Center, 00-716 Warsaw,
Bartycka 18, POLAND, sikora@camk.edu.pl, moderski@camk.edu.pl}
\author{Jill Bechtold}
\affil{Steward Observatory, University of Arizona, 933 North Cherry
Avenue, Tucson, AZ 85721, USA, jill@as.arizona.edu}
\lefthead{B{\l}a\.zejowski et al.}
\righthead{PKS 1127-145}

\begin{abstract}

We model the broad-band spectral energy distribution of the innermost
``core-jet'' region of the redshift z=1.187 quasar PKS 1127-145.  We
propose a scenario where the high energy photons are produced via the
Compton scattering of thermal IR radiation by the relativistic
particles in a parsec-scale jet.  
 The  high energy spectrum, together with the observed radio variability
and superluminal expansion, suggest that PKS 1127-145 may be a blazar,
despite the fact that its optical/UV component is likely dominated by
thermal radiation from an accretion disk.  The relation of PKS
1127-145 to MeV - blazars is discussed.

\end{abstract}
\keywords{gamma rays: theory --- infrared: galaxies --- quasars: individual (PKS 1127-145) --- X-rays: general}


\section{Introduction}

X-ray emission from  the central regions of Active Galactic Nuclei (AGN)  
is associated with accretion flows or jets. The latter dominate the
observed X-ray fluxes in blazars, the objects with powerful, relativistic 
jets oriented close to the line of sight. The X-ray spectra of these
objects extend to high energy $\gamma$-rays and  origin of both  
depends on the total luminosity of a source (e.g. Fossati et al. 1998). 
In the least luminous sources, called High Energy Peaked BL Lacs, 
the X-rays are produced by synchrotron emission, while the $\gamma$-rays -- 
presumably by  Comptonization of the synchrotron radiation (SSC).
In intermediate luminosity objects -- Low Energy Peaked BL Lacs -- 
both X-rays and $\gamma$-rays are likely produced by the SSC process.
Finally, in the most luminous sources --
Flat Spectrum Radio Quasars (FSRQ) -- the high energy spectra are dominated
by so-called external radiation Comptonization 
(ERC) process, but with a possible contribution to the soft/mid X-ray bands
from the SSC radiation. Seed photons for the ERC process may come
directly from the accretion disk (Dermer \& Schlickeiser 1993), from
the broad line region (BLR) (Sikora, Begelman \& Rees 1994), and from
an optically thick molecular torus  in the form of thermal IR 
radiation 
from hot dust
(B{\l}a{\.z}ejowski et al. 2000). The luminosity of FSRQ detected with the  
Compton Gamma Ray Observatory ({\it CGRO}) is usually dominated
by GeV radiation. However, there is a subclass of FSRQ with their
luminosity peaked in the MeV band.  Hereafter we call the former
'GeV-blazars', and the latter 'MeV-blazars'.  A possible explanation
for this division could be related to the location of the radiation
site in the jet.  As discussed by Sikora et al. (2002), in
GeV-blazars radiation is produced closer to the central black hole,
while in MeV-blazars, the active region is significantly
 farther from the center (see Sect. 3 for details).
Currently, there are about a dozen known MeV-blazars  or candidates 
for such a class;
unfortunately any searches for them are limited by the low sensitivity of
$\gamma$-ray detectors in the $1 - 30$ MeV band, where their spectra
often have a peak.

PKS 1127-145 is a radio loud quasar at redshift $z=$1.187 with a radio
spectrum peaked around 1~GHz, characteristic of the GigaHertz-Peaked
(GPS) radio-sources (Stanghellini et al. 1998; O'Dea 1998). {\it
Chandra} observations have revealed a large scale (at least $\sim$300
kpc) X-ray jet associated with weak radio emission (Siemiginowska et
al. 2002). The extended radio structure is much weaker (nearly 3
orders of magnitude) than the core emission. This value is
significantly lower in comparison to the typical FRI and FRII sources
(Kellerman \& Owen 1988).  VLBI observations show a double component
with a parsec scale jet on one side (Wehrle et al. 1992; Stanghellini
et al. 1998). The core emission dominates also in X-rays and the total
kpc-scale X-ray jet emission is $\sim$ 60 times fainter than the core
emission (Siemiginowska et al. 2002).  The strong X-ray core has
also been detected by BeppoSax (Giommi et al. 2002), 
with the spectrum
extending up to $\sim 300$ keV. Finally, a very likely counterpart to
that core is an EGRET source 2 EG J1134-1515 (Thompson et al. 1995).

We propose that the high energy emission originating in the central
parsec-scale region of the quasar PKS~1127-145 can be described in
terms of the blazar phenomenon. 
This can be justified by the following:

\begin{itemize}

\item  as is the case for many $\gamma$-ray detected blazars,  but
contrary to
steep spectrum radio quasars (SSRQ), the X-ray and $\gamma$-ray
luminosities of PKS~1127-145 exceed the optical/UV luminosity by a
very large factor (for X-rays in SSRQ see, e.g., Sanders, et
al. 1989; Laor et al. 1997);

\item the X-ray spectrum  of PKS~1127-145 is much harder than in SSRQ 
($\alpha_x \simeq 0.3-0.5$ vs. $0.6 - 0.8$) (e.g., Reeves \& Turner 2000);

\item  in contrast to SSRQ, the luminosity is peaked in the $\gamma$-ray band 
(Thompson et al. 1995);

\item the radio spectrum is flat and variable (Wehrle et al. 1992),
and superluminal motion is observed (Jorstad et al. 2001).

\end{itemize}

At the same time, unlike a typical blazar, PKS 1127-145 has a very low
optical polarization (Impey and Tapia 1990) and optical variability
(Bozyan et al. 1990). This suggests a thermal origin of the optical/UV
radiation. Such co-existence of the strong, non-thermal $\gamma$-ray
component with a prominent thermal UV bump is a characteristic of the
MeV-blazars.  We show that the jet  emission model developed by
Sikora et al. (2002) for MeV-blazars can explain these spectral properties
and make prediction that PKS 1127-145 can be variable  in X-rays
and $\gamma$-rays on the time scale 
of the order of a month.

\section{The observations}
\noindent

We obtained the multiwavelength spectrum of PKS 1127-145 shown in
Figure 1. from the literature.  The radio data (Steppe et al. (1995),
Geldzahler \& Witzel (1981), Kuhr et al. (1981), Shimmins \& Wall
(1973), Wright \& Otrupcek (1990), Griffith et al. (1994), Condon et
al. (1998), Large et al. (1981), Douglas, Bash \& Bozyan (1996)) were
extracted from the NED archives. The JHK photometry points are from
2MASS IRSA
catalog{\footnote{http://irsa.ipac.caltech.edu/index.html}}.  The
EGRET data are taken from Thompson et al. (1995).  The X-ray data are
from {\it Chandra} X-ray Observatory  (Bechtold et al. 2001, see
discussion below) and from BeppoSax (Giommi et al. 2002).

{\it Chandra} ACIS-S observations of PKS 1127-145 (conducted on
2000-05-28) provide constraints on the jet model parameters. We
analyzed the ACIS-S data in CIAO and found the best fit parameters for
the assumed absorbed power law model fit to the Chandra X-ray spectrum
(0.3-7~keV; see analysis details in Bechtold et al. 2001, and
Siemiginowska et al. 2002): photon index $\Gamma$ = 1.29$\pm$0.05,
absorbing equivalent Hydrogen column N$^{z_{abs}=0}_H$= 5.66$\pm 0.07
\times 10^{20}$~atoms~cm $^{-2}$ (in excess to the Galactic column of
3.8 $\times 10^{20}$). This model gives a flux (2-10keV) = $5.35
\times 10^{-12}$ ergs cm$^{-2}$ s$ ^{-1}$ which corresponds the X-ray
luminosity of $4.9 \times 10^{46}$ ergs s$^{-1}$.  Such a value of
$\Gamma$ and the X-ray luminosity are consistent with the X-ray
emission coming from the jet and a possible relation of this source to
the MeV-blazars which we consider below.   This is also strongly
supported by a detection of PKS 1127-145 at hundreds keV with BeppoSax
and at hundreds MeV with EGRET.

\section{The model}

Here we apply the {\it shock-in-jet} model scenario (e.g. Sikora et
al. 2001) to the X-ray emission of PKS~1127-145.  The emission
processes contributing to the X-ray and gamma-ray band are related to
the Inverse Compton scattering of the soft photons by the relativistic particles
in the jet. In the Synchrotron Self-Compton (SSC) process the synchrotron
photons emitted by the jet relativistic particles are Compton
scattered by the same population of particles. In the External Radiation
Compton (ERC) process the seed photons are located outside the jet.

We assume that due to collisions of two inhomogeneities propagating
down the jet with different velocities, the shock is formed at a 
distance $r_0$ from the jet apex and terminated at $2r_0$. The shock
propagates down the conical jet with bulk Lorentz factor
$\Gamma_{b}$. In our model
the high energy radiation
is produced via the Compton process 
directly by accelerated electrons/positrons.  The particles are assumed
to stay in the acceleration zone for a much shorter period of time
than the life-time of the shock and their acceleration/injection is
approximated by a power-law function, $Q
\propto \gamma^{-p}$, with the energy ranging from $\gamma_{min}$ to
$\gamma_{max}$.

In the External-Radiation-Compton scenario
the X-rays are produced by the Compton scattering off the electrons in
the slow cooling regime, and the gamma-rays by the Compton
scattering off the electrons in the fast cooling regime (see Sikora et
al. 2002 for details). The electron energy dividing both regimes is
given by equality of the electron cooling time scale, $t_{c}' =
\gamma /\vert \dot \gamma
\vert$, to the shock life-time, $t'\simeq \Delta r_{coll}/(c\Gamma_b )$,
where $\vert \dot \gamma \vert
\simeq c \sigma_T u_{ext}' \gamma^2/(m_e c^2)$, $u_{ext}'$ is the
energy density of the external radiation field, and $\Delta r_{coll}$
is the 
distance over which the collision of inhomogeneities takes
place. 
The primed quantities are as measured in the shock comoving
frame. Hence, the dividing energy is
\be 
\gamma_c= 
{{m_e} {c^2} \Gamma_{b} \over  \sigma_T \Delta r_{coll} u_{ext}'}\,. 
\ee

\noindent
The dividing energy  is imprinted in
the Compton spectrum at the frequency:
\be 
\nu_c \simeq {{\cal D}^2 {\gamma_c}^2 \nu_{ext} \over 1+z}\,, 
\ee
where $\nu_{ext}$ is the frequency of the external photons and 
${\cal D}=1 / \Gamma_{b} (1- \beta cos \theta)$ is the Doppler factor
of the radiating shocked plasma.
The spectrum around $\nu_c$ changes its slope by $\Delta \alpha _{x
\gamma} \simeq 0.5$. 

The soft photons contributing the most to the external radiation
field, $u_{ext}'$  in the AGN cores could come 
from an accretion disk, from the Broad
Emission Line Region (BEL) or from hot dust (IR) associated with an 
obscuring torus.  Because of
spectral similarities of PKS 1127-145 with the MeV-blazars we assume
that the high energy radiation in this object is produced at a few parsecs,
where $u_{ext}'$ is strongly dominated by thermal radiation of the hot dust.  
 PKS 1127-145 was not detected in the mid- and far-IR bands.
However, the measured spectral turnover in the $j$, $h$, and $k_s$ bands
suggests  a relatively strong contribution from the hot dust. 

The presence of the hot dust in AGN is generally accepted as a part of
the AGN paradigm (Urry \& Padovani 1995). 
The minimal distance of the dust from the central engine is
determined either by the maximal temperature it can survive,
\be
r_{d,min} \sim {1 \over {T_{d,max}}^2} \times 
(L_{UV} /4 \pi \sigma_{SB})^{1/2}\,,
\ee
where ${T_{d,max}} \simeq 1000-1500$K and $L_{UV}$ is the disk
luminosity, or by the inner edge of the molecular torus provided it is
larger than $r_{d,min}$ given above (see, e.g., Yi, Field, \& Blackman
1994).  In our calculations we approximate the dust distribution as
spherical and recover its minimum distance/maximum temperature from a
model fit to the observed high energy spectrum.  It should be
emphasized here, that as long as the distance of the shock region
 from the black hole  is
smaller than the distance of dust, the value of $u_{ext(dust)}'$ 
is not very
sensitive to dust geometry, e.g. whether the dust is distributed spherically or is located
in the flattened torus.  The density of the external IR radiation
field can be approximated by the formula:
\be
u_{ext(dust)} \sim {{\xi_{IR} L_{UV}} \over 4\pi c {r_{d,min}}^{2}} \times
{1 \over {1+(r/r_{d,min})^2}}\,,
\ee
where $\xi_{IR}$ is the fraction if the UV radiation converted by dust 
into thermal radiation.

We apply the model to PKS~1127-145 assuming that the parsec scale jet
is pointing at or very close to the line of sight. This can be
justified by detection of a superluminal motion in this source during
the Multi-epoch VLBA observations with $\beta_{app} \sim 19$ (Jorstad et
al. 2001). We compute the model by fitting
the minimal distance/maximum temperature of dust to the data. The
results are presented in Figure 1, which shows 
the contributing model components: synchrotron emission, SSC, 
ERC(IR), ERC(BEL) and ERC(UV).

The ERC(BEL) is computed assuming that at
$r>r_{BEL}$ luminosity of BEL drops with a distance
$\propto 1/r$, where $r_{BEL}$ is the distance at which the BEL
luminosity peaks (e.g., Peterson 1993; Kaspi 2000).
 Parameters of the ERC model are given in the figure caption.
They are obtained by matching: the Chandra X-ray spectrum, the EGRET
flux, and the high frequency radio flux. The emission at lower frequencies
(below $\sim 10^{10}$ Hz) is produced at significantly larger
distances from the jet apex (in comparison to our model) and thus, is
not taken into account in our modeling of the spectrum.

In Fig.1. we show also contribution into X-ray and $\gamma$-ray bands 
from Comptonization of direct accretion disk radiation (ERC(UV)). 
Using analytical
approximations given by Dermer \& Schlickeiser (2002), one can find that
this contribution at a distance of a few parsecs is about four orders lower
than the contribution from Comptonization of IR radiation of a hot dust.

 Since observational data in different spectral bands are not
simultaneous, the model spectral fit and its parameters should be
considered only qualitatively. However, the Chandra and BeppoSax
observations which were separated by 12 months, are both consistent
with the extension of a very hard X-ray spectrum up to MeV energies
and peaking at the level of the EGRET detection. Therefore the main
result of our model --- that X-ray emission of PKS 1127-145 is
dominated by Compton scattering of the infrared photons --- is not
sensitive to the spectral changes between the observations.

\section{Discussion}

Our model shows that the X-ray and gamma-ray spectrum of PKS~1127-145
may originate in the parsec scale jet. This emission can be described
entirely by the Comptonization of the IR radiation
with a possible contribution from ERC(BEL) at energies $>1$GeV. This
object is similar to the other MeV-blazars where the SSC
component is very weak and, therefore, the X-ray spectrum has a very
hard slope down to the lowest energies.

Another interesting feature of PKS~1127-145 is the co-existence of a
prominent UV thermal bump (as seen in radio-lobe dominated quasars),
with the extremely luminous $\gamma$-ray component. Both components
are very common for the MeV-blazars (see e.g. Tavecchio et
al. 2000). In our model this
co-existence can be associated with the distance between the shock and
the jet apex which affects the synchrotron emission.
The ratio $u_{ext}'/u_B'$ is higher for larger shock distances in MeV-blazars
than for smaller shock distances in GeV-blazars, where in the former case
the external radiation is dominated by infrared radiation of dust, while 
in the latter case it is dominated  by broad emission lines, and
$u_B' = B'^2/8 \pi$ is the magnetic energy density in the shocked plasma
in a jet.
 This results in a weaker
synchrotron spectrum in MeV-blazars than in GeV-blazars assuming that
both emit comparable bolometric luminosity.  Secondly, the entire
synchrotron component is shifted into lower frequencies in MeV-blazars
(because the average frequency of the synchrotron radiation is
$\nu_{SYN} \propto B^{'}$ and in our model magnetic field scales with
distance like $B^{'} \propto 1/r$).  Both effects diminish the
synchrotron contribution to the UV-band, thus the thermal UV-bump becomes
visible. Lower synchrotron luminosities in MeV-blazars explain  also
why in these objects the SSC component is weak ($L_{SSC} \propto
{L_{SYN}}^2$) and, therefore, its contribution to the X-ray band is
negligible (see Sikora et al. 2002).


The available EGRET data do not constrain the location of the
luminosity peak and do not provide a spectral slope for the
PKS~1127-145 $\gamma$-ray spectrum. Therefore, in order to minimize
the number of free model parameters we adopted a single-power law
electron injection function, instead a double-one assumed in modeling
of other MeV-blazars (Sikora et al. 2002). This approach leads to a
different location of the luminosity peak in our model spectrum of PKS
1127-145 ($h\nu \sim 100$MeV) than in the models of the other
MeV-blazars ($1$MeV $<h\nu < 30$MeV). Better $\gamma$-ray data are
needed in order to apply the same model to both MeV-blazars and
PKS~1127-145 and to constrain better the model parameters. However,
based on the current observations of PKS~1127-145, we conclude that it
is quite similar to other MeV-blazars.  Since in the model, 
high energy radiation is produced at a few parsecs, we can  predict
that PKS~1127-145 should be variable in the X-ray and $\gamma$-ray bands 
on a time scale of the order of a month. 
Monitoring of
the source with gamma-ray instruments such as Integral or GLAST may
provide an  opportunity to verify this prediction. 

PKS~1127-145 has been included in samples of GigaHertz Peaked Spectrum
(GPS) radio sources (Stanghellini et al. 1998). GPS sources are
considered to be at an early stage of their expansion into large scale
radio sources.  However, GPS samples are heterogenous and may contain
many blazars  (Lister 2003; Siemiginowska et al. 2003; Stanghellini
2003).
The X-ray data provide a way to identify
blazars and separate them from the ``true'' GPS sources. It is
important to understand a number of blazars in the GPS samples, so we
can test whether they are consistent with blazars being ``young'' GPS
sources, but seen along the jet, with the emission from the GPS double
radio component suppressed by the parsec scale jet emission. This
is critical to our understanding of the nature and evolution of radio
sources.

Finally, it is worth noting the similarity of the PKS 1127-145 with
other high-z X-ray luminous quasars with very hard X-ray spectra,
e.g.: 1508+5714 (Moran \& Helfand 1997); GB 1428+4217 (Fabian et
al. 1998; Fabian et al. 2001); PKS 2149-306 (Elvis et al. 2000); PMN
J0525-3343 (Fabian et al. 2000). All of them are FSRQ and with their
 peculiar spectra, similar to those of MeV-blazars,
they can represent those FSRQ which
have nuclei embedded in a very dusty environment. Such hypothesis can be further explored by detailed infrared
spectroscopy of those sources and by observations of $\gamma$-ray
spectra with GLAST which are expected to be relatively soft.

\acknowledgements

\noindent
This research is funded in part by NASA contract NAS8-39073and in part 
by NASA
through Chandra Award Number GO-01164X and GO2-3148A issued by the Chandra
X-Ray Observatory Center, which is operated by the Smithsonian Astrophysical
Observatory for and on behalf of NASA under contract NAS8-39073 
and by Polish KBN grant 5 P03D 00221.  We would like to thank
G.~Madejski and anonymous referee 
for their valuable comments which helped to improve the paper.

\noindent
 We acknowledge the  use of the NASA/IPAC Extragalactic Database (NED) 
which is operated by the Jet Propulsion Laboratory, California Institute of 
Technology, under contract with the National
            Aeronautics and Space Administration.

\clearpage


\centerline{\bf FIGURE CAPTIONS}

\figcaption{The broad-band spectrum of PKS 1127-145 and the applied model
components: SYN - synchrotron radiation, SSC - Comptonization of synchrotron
radiation, ERC(IR) - Comptonization of the dust radiation, ERC(BEL) -
Comptonization of the BEL radiation, ERC(UV) - Comptonization 
of the radiation from the accretion disk (assuming $M_{BH}=10^9 M_{\odot}$), hot dust - mono-temperature
thermal dust radiation, uv bump - UV radiation from the disk. The
observational data are obtained from archival data, NED,  Chandra
(bow-tie in the energy range $0.66 - 22$ KeV), and BeppoSAX.
The model parameters are as follows: $r_0=6.25 \times 10^{18}$cm;
$\gamma_{min}=1.0$; $\gamma_{max}=6.0 \times 10^3$; the bulk Lorentz
factor $\Gamma=10$; the half-opening angle of the jet:
$\theta_j=1/10$; the observer is located at an angle:
$\theta_{obs}=1/10$; the magnetic field scales with the distance like
$B^{'}(r)=(8.0 \times 10^{17})/r$ Gauss; the luminosity of the disk
$L_{UV}=1.0 \times 10^{47} erg/s$; maximal temperature of the dust
$T_{d,max}=800$ K; covering factor of the dust $\xi_{IR}=0.5$ (part of
the central UV radiation converted by dust into thermal IR radiation);
covering factor of the BEL clouds $\xi_{BEL}=0.08$ (part of the
central UV radiation converted by BLR clouds into BEL radiation),
minimal distance of the dust $r_{d,min}=1.84 \times 10^{19}$ cm,
$p=1.55$. The following cosmology has been used:
$\Omega_{\Lambda}=0.7$, $\Omega_{m}=0.3$, $H=66$ km s$^{-1}$
Mpc$^{-1}$.}

\newpage
\plotone{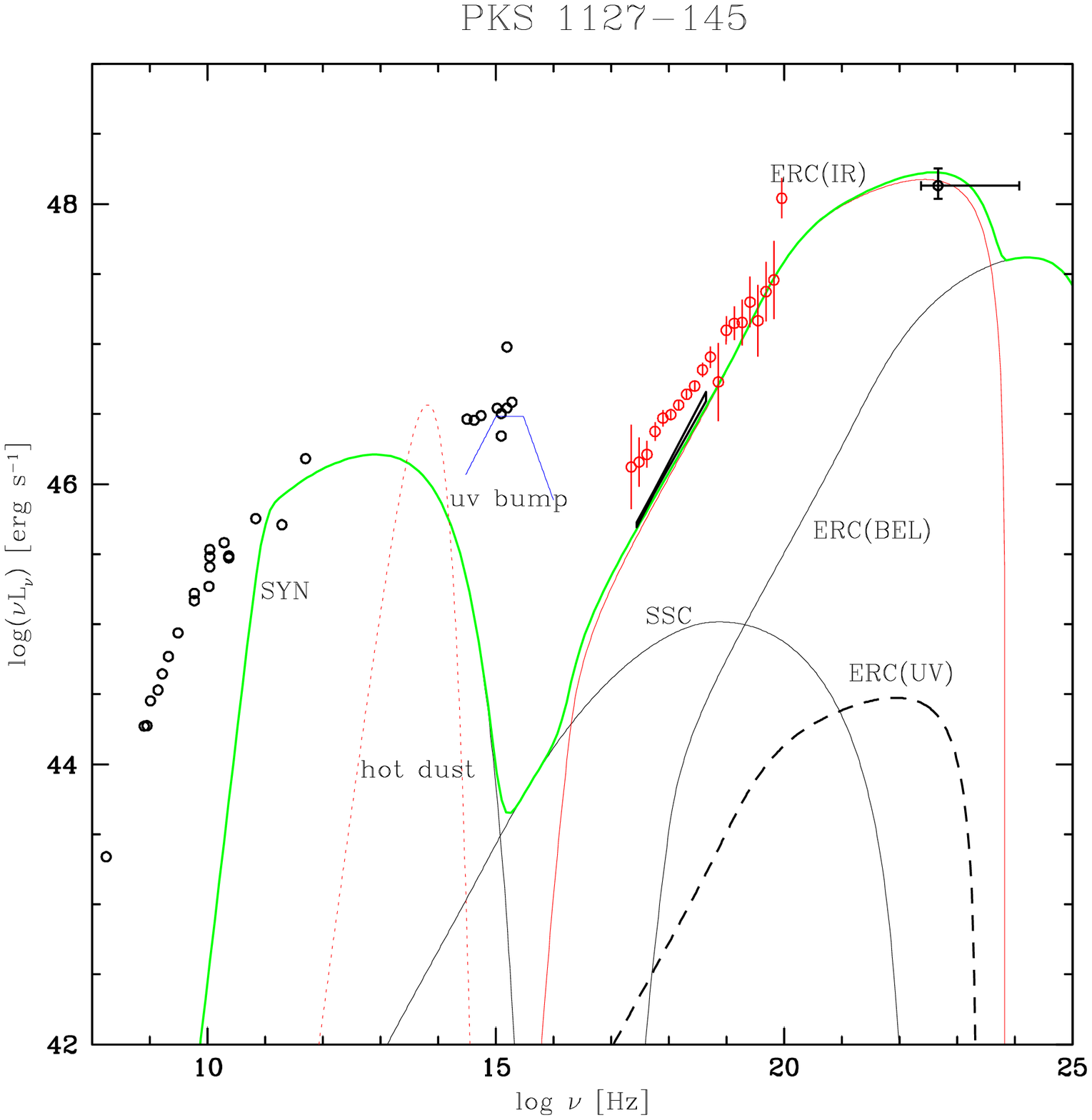}


\end{document}